\documentclass[11pt,a4paper]{article} 

\usepackage{jcappub}
\usepackage{bm}

\def\a{\alpha}

\def\d{\delta}
\def\D{\Delta}

\def\l{\lambda}

\def\M{{\cal M}}

\def\ra{\rightarrow}

\title{Fractal analysis of the large-scale stellar mass distribution in the Sloan Digital Sky Survey}

\author{Jos\'e Gaite}

\affiliation{
Applied Physics Dept.,
ETSIAE, Univ.\ Polit\'ecnica de Madrid, E-28040 Madrid, Spain}

\emailAdd{jose.gaite@upm.es}

\abstract{A novel fractal analysis of the cosmic web structure is carried 
out, employing the Sloan Digital Sky Survey, data release 7. We consider
the large-scale {\em stellar mass} distribution, unlike other 
analyses, and determine its multifractal geometry, which is compared with the geometry of 
the cosmic web generated by cosmological $N$-body simulations.
We find a good concordance, the common features being: (i)
a minimum singularity strength $\a_\mathrm{min} = 1$, which corresponds to the edge of diverging gravitational energy and differs from the adhesion model prediction; 
(ii) a ``supercluster set'' of relatively high dimension where the mass concentrates; and
(iii) a non-lacunar structure, like the one generated by the adhesion model.
}


\toccontinuoustrue

\begin{document}

\maketitle

\section{Introduction}
\label{intro}

The large scale structure of the universe can be described as a ``cosmic web,'' with characteristic though irregular geometric features that extend over lengths of tens of megaparsecs. On larger scales, the isotropy and homogeneity of the universe gradually manifest themselves, in accord with the Cosmological Principle. 
The cosmic web structure can be generated by models of the cosmic gravitational dynamics,
namely, the Zeldovich approximation and the adhesion model
\cite{Zel,GSS} and has been 
found
in galaxy surveys \cite{EJS,Ge-Hu} and cosmological $N$-body simulations
\cite{KlyShan,WG,Kof-Pog-Sh-M}.
The structure actually consists of a 
web of filaments and sheets 
of multiple sizes, 
which represent the patterns of gravitational clustering of matter.
This type of geometric structure, 
with features on ever decreasing scales, belongs in the domain of fractal 
geometry.
Of course, the geometric features of the real cosmic-web have sizes
that are bounded below by some scale determined by non-gravitational dynamics
(and above by the homogeneity scale). 

Fractal models of the universe arose from
the idea of a hierarchy of galaxy clusters that continues 
indefinitely towards the largest scales \cite{Mandel},
an idea that originated a debate about the scale of transition to homogeneity 
\cite{Cole-Pietro,Peebles,Borga,Sylos-PR,I0,Jones-RMP}.
Many fractal analyses of galaxy clustering have been motivated or
influenced by this debate, but the fractal analysis of the large-scale structure 
of the universe is interesting in its own right.
Early fractal analyses and, specifically, multifractal analyses 
comprise analyses of the distribution of galaxies 
\cite{Pietronero,Jones,Bal-Schaf} and also of the distribution of dark matter in
cosmological $N$-body simulations \cite{Valda,Colom,Yepes}.
Recent $N$-body simulations have better resolution and their  
analysis reveals new fractal aspects of the cosmic web
\cite{fhalos,I4,voids,MN,ChaCa,Bol}.  
However, the combined studies of galaxy surveys and $N$-body simulations have not led to a full description of the fractal geometry of the cosmic web
and, in particular, to a definite relation between the 
geometries of the distributions of galaxies and of dark matter. For example, 
it is not clear that the much studied power-law dependence of the galaxy-galaxy correlation  function coincides with an analogous dependence of the correlation function of
the dark matter distribution. 
It is not even clear that these two correlation functions can be directly related, because 
individual galaxies are not like dark matter particles. 

Galaxies are visible because of their baryonic content, 
and the combined dynamics of cold dark matter and baryonic gas has also 
been the object of simulations of large-scale structure formation.
The comparison between the fractal features of the baryonic gas and dark matter 
distributions in the result of one of these simulations
shows that they 
are essentially equal \cite{MN}. 
This suggests that a direct comparison of the fractal features of the
distributions of observed visible mass 
and of simulated dark matter should show good concordance.

The study of galaxy clustering by means of
correlation functions
considers galaxies
as equivalent point-like particles 
\cite{Peebles}, 
much like the particles of dark matter or gas of cosmological $N$-body simulations. 
Of course, real galaxies are not point like and their spatial 
extensions are considerable, and even larger than their visible components (even considering only the baryonic part). 
Nor are galaxies equivalent to one another. 
Galaxy catalogs provide us with their locations as point-like particles 
and also with some characteristics
that actually distinguish them 
but normally do not provide us with their masses.
The neglect of galaxy masses in the analysis of the large-scale structure
is equivalent to assigning the same mass to all galaxies, which is a questionable approximation. 
Pietronero \cite{Pietronero} already noticed the broad range of known galaxy masses and argued that it
makes the properties of the spatial mass distribution
substantially more complex, 
to such extent that a multifractal analysis is necessary,
instead of the calculation
of correlations of galaxy positions.
However, in absence of galaxy masses in the catalogs, the multifractal analysis 
that has been usually performed only
considers the number density of galaxies
\cite{Borga,Jones-RMP}.
A notable exception is the early work of Pietronero and collaborators
in which they 
calculated the masses of galaxies from the observed luminosities by
assuming a simple mass-luminosity relation
(a power law)
and thence 
carried out a proper multifractal analysis \cite{Cole-Pietro,Sylos-PR}.

That work of Pietronero {\em et al} as well as the contemporary multifractal analyses of 
other researchers that did not take galaxy masses into account were limited by the galaxy catalogs then available.
Fortunately, we have now available 
better catalogs, which contain, in particular, good estimates of stellar masses of galaxies, obtained with sophisticated methods \cite{Kauff,Blan-Ro}.
These masses
can be used to achieve a more realistic description of the distribution of 
visible mass.
In fact, a quantitative comparison between the statistical properties of the distributions of matter in galaxy surveys and
of gas or dark matter in 
$N$-body simulations requires us to take galaxy masses into account.
In contrast, the spatial extensions of galaxies are hardly relevant for 
the study of the large-scale structure.

We analyze in this work 
the galaxy distribution in the Sloan Digital Sky Survey, data release 7 (SDSS-DR7), 
employing the New York University Value-Added Galaxy Catalog (NYU-VAGC) \cite{VAGC}
and taking into account the stellar mass content of galaxies. 
We restrict ourselves to the statistical and geometric properties of the cosmic web
that can be determined by a multifractal analysis 
(for a morphological analysis of the supercluster-void network in 
SDSS-DR7, see Ref.~\cite{Tago}).
Previous studies of the distribution of SDSS galaxies in redshift space 
are mainly concerned with
the problem of the transition to homogeneity, namely, the transition from middle-scale power-law correlations to very-large-scale uniformity 
\cite{SyL-Bar,Sar-Yad,Verev,CSF}. These studies do not consider the galaxy masses  
and look for homogeneity in the number density of galaxies.
The scale of homogeneity will feature in our multifractal analysis, as a relevant parameter, but we focus on the properties of the multifractal spectrum and its comparison with the multifractal spectrum found in 
$N$-body simulations of the Lambda cold dark matter (LCDM) model, with or without gas.
At any rate, we will try to compare our results with previous results, especially, with the results of Verevkin {\em et al} \cite{Verev}
and Chac\'on-Cardona {\em et al} \cite{CSF}, who also study 
the SDSS-DR7. 
We also compare our analysis with multifractal analyses of older catalogs.

To summarize this work, we first describe the SDSS data employed
and the definition of volume-limited samples (Sect.~\ref{DR7}). 
Next, we describe 
the details of our method of multifractal analysis, including examples of 
its application to $N$-body simulations (Sect.~\ref{MF}). The results of the analysis of 
three volume-limited samples of the SDSS, 
especially, the stellar-mass distribution multifractal spectrum, 
are contained in Sect.~\ref{results},
including a comparison 
with the results
of cosmological $N$-body simulations.
Finally, we present our conclusions and discuss them in Sect.~\ref{discuss}.

\section{Samples of galaxies from the SDSS DR7}
\label{DR7}

The Sloan Digital Sky Survey in its seventh data release \cite{2009ApJS..182..543A} 
covers one quarter of the sky and has information about galaxies, quasars and stars. The 
galaxy data have been improved 
and included in the The New York U.\ Value-Added Galaxy Catalog (NYU-VAGC)
by the research group of M.R.\ Blanton and D.W.\ Hogg
\cite{VAGC}.
Two cuts in apparent magnitude are made in the Petrosian r spectral band, 
located at  $6165$  \AA: the upper cut, 
which has to be present in every survey, 
indicates the faintest objects that can be detected, while 
the lower cut is made
to prevent contamination by very bright objects.
Following Ref.~\cite{Tago}, we take the lower cut at apparent magnitude 
$m_\mathrm{r}=12.5$ and use the range $12.5 < m_\mathrm{r} < 17.77$.

The SDSS-DR7 galaxy redshifts extend from nearly null redshift to $z \simeq 0.4$.
We choose for our initial sample the following redshift limits:
$z > 0.001$, to exclude galaxies with considerable peculiar velocities, 
and $z < 0.1$, because we do not need the sample to be deep. Indeed, 
SDSS galaxies with $z>0.1$ are more luminous and their number densities 
are considerably smaller, as shown by the luminosity function of 
SDSS galaxies 
\cite[Fig.~8]{Blanton_LF}. 
Therefore, the galaxies with $z>0.1$ of the SDSS-DR7 are
less suitable for obtaining information on the smaller scales.
We further restrict our initial sample to the SDSS main angular area (see Sect.~\ref{coor}).
The number of galaxies in the sample is 305\hspace{1pt}854.

\subsection{Volume-limited samples of galaxies}

As in any redshift survey with limits in apparent magnitude, 
the mean number density of SDSS-DR7 galaxies decreases with redshift.
It is necessary to construct {\em volume-limited} subsamples of the full sample
to correct this radial selection effect 
\cite{Sylos-PR,Jones-RMP}. 
Unlike in Refs.~\cite{SyL-Bar,Sar-Yad,Verev,CSF}, in  which 
volume-limited subsamples of galaxies are defined by their ranges of absolute magnitudes, 
here they are defined by redshift ranges, which determine 
the corresponding absolute magnitude ranges.  
The cuts in absolute magnitude are given by the expression of the absolute magnitude 
in terms of apparent magnitude and redshift
\cite{Peebles}:
\begin{equation}
M_\mathrm{abs}= m_\mathrm{r} - 5\log_{10} R(z)-25-K(z),
\label{MVL}
\end{equation} 
where $m_\mathrm{r}$ is the apparent magnitude 
in the Petrosian r band, $R$ is the
luminosity distance in Mpc, and $K(z)$ is the k-correction for the SDSS r band.
In Refs.~\cite{SyL-Bar,Verev,CSF}, two different approximations for the
calculation of the k-correction have been employed. Here we employ the 
approximation of Chilingarian et al \cite{Chili}, which is appropriate in our case. 

Since we construct volume-limited (VL) samples by specifying their ranges of redshift
and the redshifts are associated to individual galaxies, 
it is easy to analyze how a VL sample changes when $R(z)$ 
or $K(z)$ change, for example, after a change of the cosmological model.
In this regard, we assume a standard LCDM cosmology, with 
$\Omega_\mathrm{m}=0.3$, $\Omega_{\Lambda}=0.7$, but we have 
checked that changes in the parameters within reasonable ranges 
do not alter the results.
For the Hubble constant, we may think of the choice $h=1$, 
but our VL samples, in terms of ranges of $z$, are 
independent of $h$, because their construction only involves 
ratios of pairs of values of $R(z)$. 
In consonance, distances will be expressed in Mpc/$h$.

Before deciding the VL samples to employ, it is convenient to 
explain our choice of coordinates, with the angular selection, and also consider 
the requirements of our 
multifractal analysis. The selection of our VL samples is described in 
Sect.~\ref{select}.

\subsection{Choice of coordinates}
\label{coor}

Let us consider first the angular coordinates for fixed radial distance. The most convenient angular coordinate systems are orthogonal systems such that they preserve the area, that is to say, such that the element of area is
just the product of the line elements along the two coordinates, 
like in Cartesian coordinates.
This type of coordinates is common in geography \cite{mathworld} and have 
been employed for multifractal analysis of $N$-body simulated halos 
in Ref.~\cite{Bol}. 
Equal area coordinates can be defined in terms of angular spherical coordinates: 
one coordinate is just the longitude and the other is the sine of the latitude.
The equatorial coordinates, namely, right ascension $\a$ and declination $\d$, are indeed 
spherical coordinates and could be used for this purpose. However, 
the SDSS imaging camera scans the sky in {\em strips} along 
particular great circles,
so that the appropriate coordinate system, called the {\em survey coordinate 
system} in Ref.~\cite{Stough}, is a different system of angular spherical coordinates,
with poles at $\a=95^\circ$, $\d=0^\circ$, and $\a=275^\circ$, $\d=0^\circ$.
The latitude and longitude measured from these poles are called $\l$ and 
$\eta$, respectively. 
In Ref.~\cite{Stough}, the origin of $\eta$ is set to the point 
$\a=185^\circ$, $\d=32.5^\circ$ but
we have set it to $\a=185^\circ$, $\d=0^\circ$, 
that is to say, the middle point of the semicircle $\d=0^\circ$, 
$\a \in [95^\circ,275^\circ]$. 
Therefore, our equal-area coordinates are $sl=\sin\l$ and 
$f=\eta+32.5^\circ(\pi/180^\circ)$ (in radians).
They are related to $\a$ and $\d$ by the transformations
{\setlength\arraycolsep{2pt}
\begin{eqnarray*}
sl &=& \sin(\a-185\pi/180)\,\cos\d, \\
\tan f &=& \tan\d/\!\cos(\a-185\pi/180)
\end{eqnarray*}}%
($\a$ and $\d$ must be expressed in radians).

\begin{figure}
\centering{
\includegraphics[width=7.2cm]{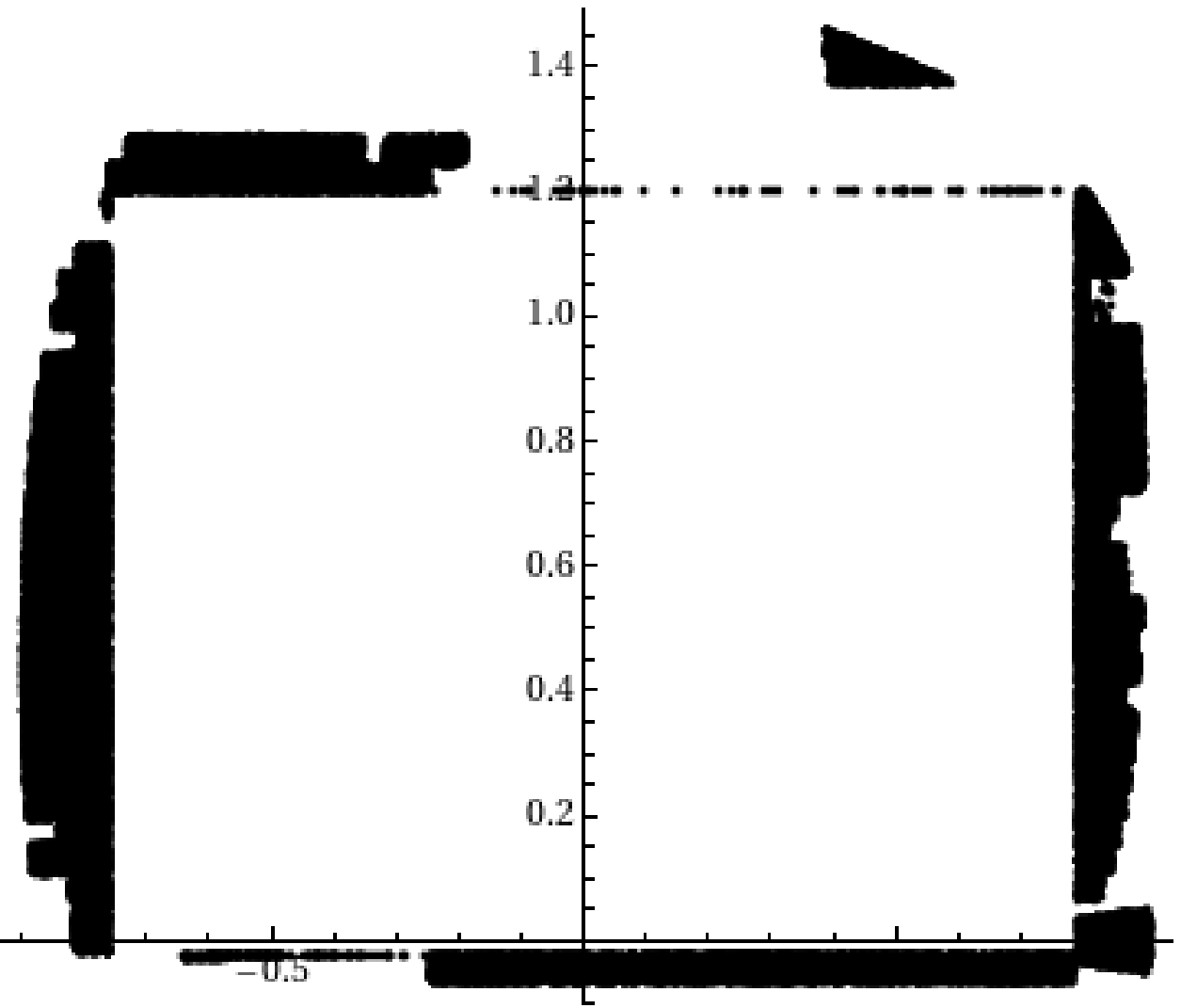}
\hspace{3mm}
\includegraphics[width=7.2cm]{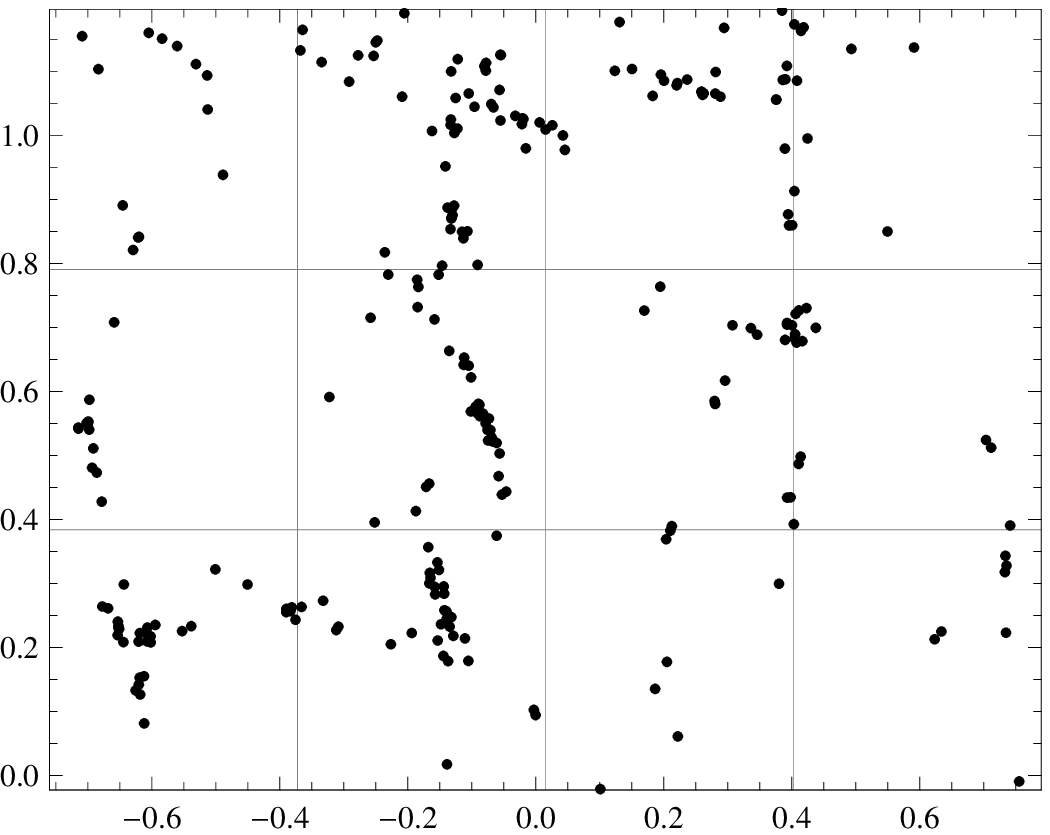}
}
\caption{Our system of equal-area angular coordinates $sl$ and 
$f$ and the selected rectangle $[-0.76,0.79] \times [−0.02,1.20]$: (Left)
The small portion of SDSS-DR7 main-area galaxies left out of the rectangle;
(right) radial slice (0.01 $< z <$ 0.0107) of the galaxy distribution in redshift space, 
showing the cosmic web structure.}
\label{slice}
\end{figure}

With this choice of coordinates, we can form a regular region, namely, 
a rectangle, such that it covers most of the SDSS main area. 
The rectangle is the product of the intervals $-0.7604<sl<0.7934$ and 
$-0.02269<f<1.1996$, covering a total solid angle that is the 
product of the respective ranges of $sl$ and $f$, 
namely, $\Omega=1.554 \cdot 1.222=1.899$ steradians.%
\footnote{The ``rectangle'' so defined is not a {\em spherical 
rectangle}, because the two opposite sides with constant $sl$, that is to say,
constant latitude $\l$, are not great circles and are therefore 
curved in the intrinsic geometry of the spherical surface.}
This rectangle is displayed in Fig.~\ref{slice}, where the fractions of 
area do represent fractions of solid angle and hence of galaxy number, 
because $sl$ and $f$ are equal-area coordinates.

For the radial coordinate $r$, the natural definition is the comoving distance that corresponds to the assumed values of 
the cosmological parameters. The relation between $r$ and the luminosity distance 
in Eq.~(\ref{MVL}) is $R = (1+z)\,r$. 

\section{Multifractal analysis}
\label{MF}

Let us assume that some mass is distributed in a region of space.
The multifractal analysis of this distribution can be carried out in two different ways:
either by using a lattice of cells (boxes) covering the region (a method called by 
Falconer \cite{Falcon} ``coarse multifractal analysis'')
or by using point-centered spheres, where the points
span the support of the distribution. 
Harte \cite{Harte} compares both methods. 
For a distribution of equal-mass particles, the calculation of 
the two-point correlation function is equivalent to the calculation of the 
point-centered second multi\-fractal moment.
In fact, the calculation of point-centered statistical moments, or just the second moment, 
in the form of conditional density, 
has been the main method of fractal analysis of the SDSS galaxy distribution
\cite{SyL-Bar,Sar-Yad,Verev,CSF}. 
However, a lattice multifractal analysis is more practical to cope with 
a large amount of data and avoids the problem of choosing a maximum radius for
point-centered spheres. 
In the analysis of 
distributions of equal-mass particles, 
this method boils down to an
elaboration of counts-in-cells statistics; but it is also practical 
for unequal-mass particles. We explain it now.

Given the lattice, 
fractional statistical moments are defined as 
\begin{equation}
\M_q = \sum_i \left(\frac{m_i}{M}\right)^{q}, 
\label{Mq}
\end{equation}
where the index $i$ runs over the set of non-empty cells, 
$m_i$ is the total mass of the
particles in the cell $i$, while $M= \sum_i m_i$ is the total mass of all the particles, and $q \in \mathbb{R}$.
Some distinguished integral moments are:
$\M_0$, which is just the number of non-empty cells; $\M_1$, normalized to one;
and $\M_2$, which is related to the two-point correlation function.

In a regular distribution, with a well-defined density everywhere, 
if we take a sufficiently fine mesh, then 
the mass contained in any cell is proportional to the cell volume $v$.
Therefore, $\M_q \sim v^{q-1}$.  
This does not apply to singular distributions.%
\footnote{A continuous mass distribution with a well-defined density
is said to be {\em absolutely continuous}. Although this property
may seem natural, the standard methods of randomly generating continuous 
mass distributions produce {\em strictly singular} distributions, namely, distributions with no positive finite density anywhere \cite{Monti}.}
But the singularities of a distribution can be such that the $q$-moments are non-trivial
power laws of $v$ in the $v \ra 0$ limit.
So one can define, for any distribution, the exponents
\begin{equation}
\tau(q) = 3\lim_{v\ra 0}\frac{\log \M_q}{\log v}\,,
\label{tauq}
\end{equation}
provided that the limit exists (for every $q$).
Such a distribution is called multifractal. 
For a regular distribution, $\tau(q) = 3(q-1)$, whereas for 
singular distributions the exponents are non-trivial.
Of course, the numerical evaluation of the limit in Eq.~(\ref{tauq}) is not
feasible and one must be satisfied with finding a constant value of the
quotient for sufficiently small $v$, that is to say, 
in a range of negative values of $\log v$ (a range of small scales).  
In fact, the exponent is normally defined as
the slope of the graph of $\log \M_q$ versus $\log v$, and its value is found by numerically fitting that slope.

The standard lattice in multifractal analysis is the Euclidean rectangular and even cubical 
lattice \cite{Falcon,Harte}. In fact, a cubical lattice is perfectly adapted
to the analysis of $N$-body simulations. 
However, volume-limited galaxy samples are defined in spherical sectors, which
makes such lattices inadequate, because they lead to a loss of data. 
It is preferable to define a rectangular lattice in the coordinates 
adapted to spherical sectors that have been defined in Sect.~\ref{coor}
(which is not a rectangular lattice in Euclidean space). 
In addition, we require that the cells have identical volume.
This can be achieved by dividing the angular-coordinate rectangle, given by 
the intervals of $sl$ and $f$, 
into equal area sub-rectangles (like in Ref.~\cite{Bol}),
and by also splitting the range of $r$ into intervals with 
constant $\D(r^3)=(r+\D r)^3-r^3$. Such a lattice is not unique and we will further 
require that the resulting cells are reasonably regular, 
with aspect ratios not very different from one (Sect.~\ref{coarse}).

Besides the moment exponents $\tau(q)$, a multifractal is also characterized by 
its {\em local} dimensions. 
The local dimension $\a$ at the point $\bm{x}$ 
is the exponent of mass growth from that point outwards, that is to say, $m(\bm{x},r) \sim
r^{\a(\bm{x})}$, where $m(\bm{x},r)$ is the mass in a ball or box of linear size $r$ 
centered on $\bm{x}$.
The local dimension measures the ``strength'' of the singularity: the smaller is $\a$, 
the more divergent is the density at $\bm{x}$ and 
the stronger is the singularity.
Actually, singularities correspond to $\a < 3$, 
that is to say, to a divergent density, 
whereas points with $\a > 3$ have vanishing density (if they exist).  
Every set of points with a given local
dimension $\a$ constitutes a fractal set with a (Hausdorff) dimension that depends on
$\a$, denoted by $f(\a)$.  In terms of $\tau(q)$, 
the spectrum of local dimensions is given by
\begin{equation}
\a(q)= \tau'(q)\,,\quad q \in \mathbb{R}\,,
\label{aq}
\end{equation}
and the spectrum of fractal dimensions $f(\a)$ is given by the Legendre
transform
\begin{equation}
f(\a) = q\,\a - \tau(q)\,.
\label{fa}
\end{equation}
Standard self-similar multifractals have a typical spectrum of fractal dimensions that
spans an interval $[\a_{\mathrm{min}},\a_{\mathrm{max}}]$, is concave (from
below), and fulfills $f(\a) \leq \a$ \cite{Falcon,Harte}.  
Furthermore, the equality $f(\a)=\a$ is
reached at one point, with $\a<3$ and such that $q=1$ in Eq.~(\ref{fa}) [notice that
Eq.~(\ref{tauq}) gives $\tau(1)=0$].  The corresponding set of singularities
contains the bulk of the mass and is called the ``mass concentrate.''

As a complement to the multifractal spectrum $f(\a)$, it is useful to define
the spectrum of R\'enyi dimensions 
\begin{equation}
D_q= \frac{\tau(q)}{q-1}\,,
\label{Dq}
\end{equation}
because they have an information-theoretic meaning \cite{Harte,Renyi}. Indeed, 
they express the power-law behavior
of the R\'enyi $q$-entropies 
of the coarse distribution in the limit of vanishing coarse-graining volume, 
$v \ra 0$.
The dimension of the mass
concentrate $\a_1 = f(\a_1) = D_1$ is associated to the ordinary entropy 
and is also called the entropy dimension.
$D_0=-\tau(0)$ is the box-counting dimension of the support of the distribution
(let us recall that the support of a mass distribution is the smallest {\em closed} set 
that contains all the mass \cite{Falcon}). 
$D_0$ also coincides with the maximum value of $f(\a)$.
$D_2 = \tau(2)$ is the correlation dimension.  
For a regular distribution, $\tau(q) = 3(q-1)$ and $D_q=\a=f(\a)=3$.
For a uniform fractal (a {\em unifractal} or {\em monofractal}), 
$\a$, $f(\a)$ and $D_q$
are also constant but $D_q=\a=f(\a)<3$.
In general, $D_q$ is a non-increasing function of $q$.

As said above, the convergence to the limit in Eq.~(\ref{tauq}) must take
place in a range of small values of $v$. Naturally, $v$ must be small in
comparison to the homogeneity volume $v_0$, which is the smallest volume
such that the mass fluctuations in it are small and approximately
Gaussian (assuming that homogeneity holds on sufficiently large scales). 
For cell sizes $v$ close to $v_0$ or larger, the fluctuations tend to 
vanish and 
$\M_q \approx v^{q-1}$.
This relation is an asymptotic equality, 
provided that the total sample volume is normalized to one. 
On account of it, 
we define, for a given cell size $v$, the {\em coarse
exponent} as
\begin{equation}
\tau(q) = 3\frac{\log (\M_q/v_0^{q-1})}{\log (v/v_0)}\,.
\label{ctauq}
\end{equation}
In this fraction, both the numerator and denominator vanish
when $v$ approaches $v_0$ from below 
(the former approximately and the latter exactly).
Their quotient tends to $\tau(q) = 3(q-1)$, namely, the form of $\tau$ for regular distributions. 
The coarse exponent (\ref{ctauq}) depends on both $v$ and $v_0$ but must 
become independent of $v_0$ when $v \ll v_0$, 
provided that the limit $v \ra 0$ exists.
However, the dependence on $v_0$ 
must not be ignored, because the rate of convergence to the limit does 
depend on the value of $v_0$. In particular,
if one sets $v_0$ to one, namely, the total sample volume, 
and this volume greatly exceeds the homogeneity volume, 
the coarse exponents can be so inaccurate that no
convergence can be observed and it is not possible to speak of a scaling limit. 
In other words, if $v_0$ is not set correctly,
the available range of $v$ may not be long enough for us to obtain reliable values
of the functions $\tau(q)$ and $f(\a)$.  
We discuss the choice of $v_0$ in Sect.~\ref{v0}.

When the cell volume $v$ 
reaches $v_0$, 
each cell can be considered as an independent 
realization of the stochastic process that generates the cosmic web structure. 
The multifractal spectrum $f(\a)$ measures the probability of 
finding mass concentrations of strength $\a$ in a realization.
This probability is estimated, in a lattice with small $v$,  
as the number of cells with 
strength $\a$ 
divided by the number of non-empty cells, 
approximately, $v^{-f(\a)/3}/v^{-D_0/3} = v^{[-f(\a)+D_0]/3}$.
This probability is maximal and close to one when 
$f(\a)$ takes its maximum value $D_0$, because most 
non-empty cells have the corresponding value of $\a$.
On the contrary, the probability is minimal for 
$\a_{\mathrm{min}}$ or $\a_{\mathrm{max}}$, because they normally occur 
only once.
If the sample only occupies the homogeneity volume ($v_0=1$), 
then we have just one realization and 
$f(\a_{\mathrm{min}})=f(\a_{\mathrm{max}})=0$, so that the cells
with $\a_{\mathrm{min}}$ or $\a_{\mathrm{max}}$ occur just once. 
A larger sample ($v_0<1$)  
contains more than one realization of the stochastic process.
If $\a_{\mathrm{min}}$ and $\a_{\mathrm{max}}$ still occur only once, 
then there is less than one cell
with $\a_{\mathrm{min}}$ or $\a_{\mathrm{max}}$
per realization and $f(\a_{\mathrm{min}})=f(\a_{\mathrm{max}}) <0$; that is to say, 
there are {\em negative fractal dimensions}.
This anomaly 
has been discussed by Mandelbrot \cite{Mandel2}. 
It is linked to the use of coarse multifractal analysis, 
to the extent that any dependence on $v_0$ must disappear as $v\ra 0$.
Indeed, when $f(\a) < 0$, the number of cells with strength $\a$ {\em diminishes} as 
$v\ra 0$. In the limit, 
any set of singularities of strength $\a$ with $f(\a) < 0$ is {\em almost surely} empty.
Therefore, we can discard this part of the spectrum.

\subsection{Errors in multifractal analysis}
\label{errors}

In coarse multifractal analysis, 
the computation of the coarse exponents (\ref{ctauq}) is subject to errors that increase 
for small cell volume $v$, with the consequent limitation of the available scale range.
%
The computation of $\M_q$ by Eq.~(\ref{Mq}) is subject to errors because 
the mass $m_i$ in each cell is uncertain. This uncertainty is 
due to the uncertainty in the positions
and extensions of galaxies
combined with 
the uncertainty in the galaxy masses, namely, in the available stellar mass estimates of the galaxies. Although the latter type of error does not exist 
if one assumes that the galaxy masses are equal, 
this assumption is intrinsically much more erroneous. 
Adequate methods of error estimation are explained 
in Sects.~\ref{MF-Bolshoi} and \ref{MFspec}, but it is convenient to consider 
before some generalities. 


The major cause of limitation in the available scale range for a sample
lies in the discrete nature of the sample: 
if there are many galaxies in a cell, 
the statistical uncertainties in positions or masses tend to compensate each other, 
whereas the uncertainty is largest 
in the cells with the smallest number of particles.
%
The effect on the coarse exponent $\tau(q)$
depends on the value of $q$.
The form of $\M_q$ in Eq.~(\ref{Mq}) shows that, 
for $q>0$, the errors in the larger values of $m_i$ are more important, whereas, for $q<0$, the errors in the smaller values of $m_i$ are more important. 
In consequence, the values of 
$\tau(q)$ for $q<0$, and hence the values of $f(\a)$ for $\a > 3$ are more affected by discretization errors and are therefore more uncertain. In general, 
we can say that the geometric features of voids 
(zones with $\a > 3$) are more difficult to establish, 
because voids are usually undersampled.  
Moreover, the uncertainty of $\M_q$ for $q<0$ increases with decreasing $v$.
The values of $\M_q$ for $q>0$ are less uncertain for a given $v$, but their
uncertainty also increases with decreasing $v$, as the discretization 
errors grow. In general, the geometry of both clusters and voids
is only discernible when they contain sufficient numbers of galaxies. 

In the case of equal masses, the variable that rules the 
discretization errors is the mean number density $n$ of the sample. 
The discretization length
$n^{-1/3}$ (length of the cube with one particle on average) is the 
overall scale for the onset of discretization errors (the smaller scales
can be said to belong to the ``shot-noise regime'' \cite{I0}). 
The structure of cosmic voids, in particular,
is lost on scales smaller than the
discretization length, while the structure of clusters persists 
on somewhat smaller scales \cite{I4,voids,MN}. In fact, it is easy to
see that singularities of strength $\a$ are sampled down to the length scale 
$v_0^{1/3-1/\a} n^{-1/\a}$, where $v_0$ is the homogeneity volume. 
This criterion is only valid for equal masses,
but it could be regarded as a heuristic rule for unequal masses.
For galaxies in a certain mass range, 
a higher number density obviously diminishes the discretization errors, but it is not easy to compare the errors for different mass ranges and how they 
affect different ranges of $\a$.
We shall see that low-luminosity and therefore low-mass volume-limited samples 
are generally preferable over the full range of $\a$ (Sect.~\ref{MFspec}).

In regard to the actual evaluation of errors, we can estimate the errors 
of the moments $\M_q$ for a given $v$, in terms of the 
errors in positions and masses (Sects.~\ref{MF-Bolshoi} and \ref{MFspec}),
but this does not really tell us how reliably 
the coarse exponents $\tau(q)$ approach their limit values for $v\ra 0$. 
In fact, the approach to the $v\ra 0$ limit is easily demonstrated by 
checking the numerical convergence of coarse exponents.
However, convergence
can only take place before discretization errors take hold, and this 
happens at larger values of $v$ for larger $\a$ (smaller $q$). 
The maximum scale range for convergence extends from the 
homogeneity scale $v_0^{1/3}$ 
down to the $\a$-dependent discretization scale written above, which
gives, in the equal-mass case, the scale factor
$$
\frac{v_0^{1/3}}{v_0^{1/3-1/\a} n^{-1/\a}}
= \left(v_0\,n\right)^{1/\a}.$$
It is indeed smaller for larger $\a$ (and tends to one for $\a \ra\infty$).
The crucial non-dimensional variable is $v_0\,n$, which is the number of particles in a homogeneity volume. This number is very large in recent 
$N$-body simulations and guarantees good convergence even for 
$\a \gtrsim 4$,  
as shown in Sect.~\ref{MF-Bolshoi}. 
Unfortunately, the situation is much worse
in galaxy surveys.

In summary, the estimation of errors in the coarse multifractal spectra for a given sample 
is less useful than the analysis of convergence of those spectra in the available scale range.
This range is limited by the growth of the magnitude of errors for diminishing $v$, which 
restricts the maximum value of $\a$ for which convergence holds.
The overall accuracy of our results relies on several consistency checks. First, 
the results for a given volume-limited galaxy sample must be self-consistent across 
its available scaling range, which amounts to a proof of convergence to a multifractal limit. 
Second, the results for different samples must be consistent. 
Finally, the multifractal geometry of the stellar mass thus obtained, after passing the preceding checks, must be consistent with the multifractal geometry of dark matter and gas derived from $N$-body simulations.

\subsection{Multifractal spectrum from cosmological $N$-body simulations}
\label{MF-Bolshoi}

\begin{figure}
\centering{\includegraphics[width=7.2cm]{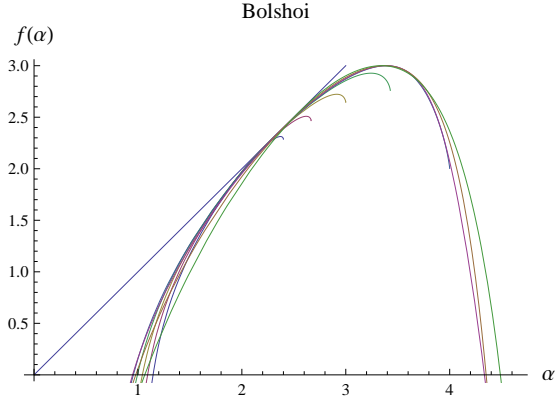}}
\caption{The multifractal spectrum of the present time dark matter distribution in the Bolshoi simulation, as
 a typical multifractal spectrum of LCDM $N$-body simulations. This 
graph clearly shows the convergence of coarse multifracta spectra.}
\label{Bolshoi_spec}
\end{figure}

The multifractal analysis of LCDM $N$-body simulations shows 
that a sufficient convergence can be achieved for all values of 
local dimension $\a$
and reveals a typical multifractal spectrum, represented in 
\cite[Fig.~5]{I4}, \cite[Fig.~2]{MN}, or \cite[Fig.~2]{Bol}. 
For example, the analysis of the Bolshoi (= Big) simulation \cite{Bol}
gives the multifractal spectrum displayed in Fig.~\ref{Bolshoi_spec}. 
Notice the good convergence of the coarse multifractal spectra, which 
correspond to the scale of 3.9 Mpc/$h$ (a fraction $2^{-6}$ of the total box length) 
and seven subsequently halved scales (a total factor of $2^7=128$). 
But only the larger scales can give the multifractal spectrum in the zone of voids ($\a>3$)
and reach the maximum value $f(\a)=3$. 
Essentially the same multifractal spectrum of Fig.~\ref{Bolshoi_spec} is found 
in other LCDM $N$-body simulations, and,
moreover, the analysis of the Mare-Nostrum simulation, which includes gas, shows that 
the same multifractal spectrum is found for the distribution of gas \cite[Fig.~2]{MN}.
The salient features of the common spectrum are the following.

First of all, it has the typical concave shape that corresponds to a self-similar 
multifractal \cite{Falcon,Harte}.
The maximum value of $f(\a)$, equal to 
the box-counting dimension of the support of the distribution $D_0$, 
is very close to 3, which is a special value.
Notice that $\M_0=v^{-1}$ 
when there are no empty cells in the lattice, yielding, according to Eq.~(\ref{ctauq}), 
$D_0=-\tau(0)=3$.
However, in the $v\ra 0$ limit, there can be empty cells
and what matters is how the number of them grows: if it does not grow at a sufficient
rate, then still $D_0=3$.
Therefore, $D_0=3$ occurs either with no voids or with a sequence 
of voids of sizes that decrease too rapidly. Indeed, what singles out 
a {\em fractal hierarchy} of voids is that it fulfills the Zipf law or 
that it follows the Pareto distribution \cite{Mandel,voids}.
The analysis of LCDM $N$-body simulations, e.g., the
Bolshoi simulation in Fig.~\ref{Bolshoi_spec}, shows that there are actually 
no empty cells, starting from $v\lesssim v_0$ and well into the scaling range, and 
the voids that arise on lower scales seem to be due to undersampling.
This suggests that the cosmic web has a {\em nonlacunar} multifractal geometry, with 
{\em no totally empty} voids \cite{voids}.%
\footnote{This conclusion refers to Mandelbrot's original definition 
of lacunarity \cite{Mandel} and does not mean that 
a non-vanishing lacunarity cannot be defined. 
Indeed, the concept of lacunarity has proved to be subtle and there are various definitions.
The notion of nonlacunar fractal, as a fractal set that is everywhere {\em dense}, 
was also introduced by Mandelbrot \cite{Mandel}.}

As regards the mass concentrate, the dimension $\a_1 = f(\a_1)$, 
given by the point of tangency to the diagonal
in the graph, seems to be about 2.4, but it cannot be 
determined precisely. 
The strongest singularities have $\a \simeq 1$. This value, 
namely, the law of mass growth $m(r) \propto r$, corresponds to 
the singular isothermal sphere profile or  
to a filament, in the fully isotropic or extremely
anisotropic cases of mass concentrations, respectively.
The strongest mass depletions have $\a \simeq 4.5$.

The best convergence of coarse multifractal spectra takes place
in the $N$-body simulations with the largest values of $v_0\,n$, 
that is to say, with the best mass resolution.
In the Bolshoi simulation, 
the homogeneity scale is one eighth of the simulation box, which yields 
an overall scale range
$\left(v_0\,n\right)^{1/3} = 256$. 
This range is reduced for $\a > 3$ but 
is nevertheless sufficient for the largest values of $\a$ present:
one can observe in 
Fig.~\ref{Bolshoi_spec}
convergence of three coarse multifractal spectra even for $\a \simeq 4.5$.
For $\a < 3$, the convergence is still more convincing.

Moreover, it is easy to see that each coarse multifractal spectrum 
has a negligible error due to errors in particle positions. 
Particle coordinates are given by {\em floating-point} numbers
with 23-bit mantissa (excluding the sign bit). The 13 most significant bits 
are preserved by the coarse graining to the smallest cell used in the Bolshoi simulation
(with 0.03 Mpc$/h$), 
leaving a 10-bit precision inside each cell. For larger cells, 
the precision is higher, of course. Therefore, 
the relative error in position inside any cell
is $< 10^{-3}$.
The relative error in mass of cells with many particles, which are more important for 
$q>0$, is proportional to the relative error in positions, and, in fact, the 
relative error in $\M_q$, for $q>0$, is practically equal to the relative error in positions. 
For $q<0$, cells with few particles can be important in the computation of $\M_q$ 
by Eq.~(\ref{Mq}). However, the sum in that formula can be expressed, 
for equal-mass particles, as a sum over number of particles per cell, so that each 
summand is multiplied by the number of cells with a definite number of particles. The 
error may alter significantly the number of particles in single cells with few particles but 
will not alter significantly each summand. In fact, the relative error in 
the number of cells with a definite number of particles is of the order of magnitude of
the relative error in position, in any case. This implies that the change in the 
coarse multifractal spectra induced by errors in particle positions is inappreciable 
in Fig.~\ref{Bolshoi_spec}.

If we consider the good concordance of multifractal spectra for the Bolshoi simulation
and other LCDM
$N$-body simulations \cite{I4,MN}, 
we can say that we have a reliable multifractal spectrum of the LCDM cosmic web.
In galaxy samples, the mass resolution is much worse and the errors are considerable, 
so we should not expect to obtain nearly as accurate a multifractal spectrum, and 
we must rather compare what we obtain to the $N$-body simulation multifractal spectrum. 
This is especially true in the zone $\a > 3$, for which the mass resolution of galaxy samples 
is hardly sufficient.

\section{Procedure and Results}
\label{results}

Here we describe in detail how we select VL samples in a few 
redshift intervals, how we determine the value of $v_0$, needed for Eq.~(\ref{ctauq}), 
and how we construct coarse lattices appropriate for the selected VL samples. Finally,
we calculate the multifractal spectra of these samples.

\subsection{Selection of volume-limited samples}
\label{select}

Previous fractal analyses of 
SDSS galaxies  
have either considered several VL samples, in ranges of consecutive absolute 
magnitudes \cite{SyL-Bar,Verev,CSF}, or focused on a particular sample
\cite{Sar-Yad}. In all cases, there has been a bias towards deep VL samples and, therefore,
high luminosities, in accord with their common goal of 
analyzing the transition to homogeneity. 

For just finding the multifractal geometry of the mass distribution, namely, 
the properties of singular mass concentrations and mass depletions, 
the relevant length scales are necessarily smaller. 
In the present multifractal analysis 
of the stellar mass of galaxies, it may seem that 
we should favor 
VL samples that
contained large fractions of the total stellar mass density. 
Indeed, such samples represent the mass concentrate set, namely,
the set of singularities that contains the bulk of the stellar mass. 
However, the bulk of the stellar mass corresponds to 
rather bright galaxies
and, hence, to moderately deep VL samples. 
Unfortunately, if we employ these samples, 
we miss information about voids, which is best obtained from 
VL samples with fainter and less massive galaxies.

 \begin{table}[h]
 \begin{center}
 \footnotesize
    	\begin{tabular}{ccccccccccccccccc}
 	\hline
	Redshift  & $r$ (Mpc/$h$) & $N$ & $V$ (Mpc$^{3}\!/h^{3}$) &$n=N/V$ & $\rho$ (M$_\odot h^{3}\!/$Mpc$^{3}$) & galaxy mass (M$_\odot$)\\
	\hline\hline
	$[0.003,0.013]$ &  $[8.99,38.9]$ & $1765$ & $3.68\cdot 10^4$ &
$0.048$ & $8.43\cdot 10^6$ & $[5.8\cdot 10^5,1.1\cdot 10^9]$\\
	\hline
	$[0.02,0.03]$ &  $[59.7,89.4]$ & $16557$ & $3.17\cdot 10^5$ &
$0.052$ & $2.65\cdot 10^8$ & $[1.7\cdot 10^6,5.5\cdot 10^{10}]$\\
	\hline
	$[0.04,0.06]$ &  $[118.9,177.5]$ & $42021$ & $2.48\cdot 10^6$ &
$0.017$ & $1.92\cdot 10^8$ & $[2.1\cdot 10^7,1.9\cdot 10^{11}]$\\
	\hline
        \end{tabular}
 \caption{Characteristics of the three volume limited samples.} 
  \label{VLStable}
 \end{center}
 \end{table}

Consequently, to construct our VL samples, 
it is convenient to take first a range of low-redshift galaxies, that is to say, 
faint galaxies, and then proceed to deeper samples. 
In fact, we can do with just three VL 
samples, chosen as follows: VLS1 with $z \in [0.003,0.013]$, VLS2 with 
$z \in [0.02,0.03]$, and VLS3 with $z \in [0.04,0.06]$ (see Table 
\ref{VLStable}).
The first one
is actually the most useful one, because not only is it useful
for the study of voids (the zone $\a > 3$) but also, 
as it turns out, for the study of clusters (the zone $\a < 3$). The reason 
is that the information on clusters can be obtained not only from the 
more massive galaxies, which are to be found mainly in clusters, 
but also from the patterns of clustering of the less massive galaxies. 
We indeed find
that the multifractal spectra obtained 
from the three VL samples in the zone $\a < 3$ are practically equal 
(Sect.~\ref{MFspec}).

\subsection{Homogeneity scale}
\label{v0}

To implement the procedure of coarse multifractal analysis 
described in Sect.~\ref{MF}, we need to 
compute the $q$-moments $\M_q$ from the set of cell masses $m_i$, and then
compute the coarse exponents $\tau(q)$, according to Eq.~(\ref{ctauq}), 
and do it for several lattices, with decreasing cell volumes $v$.
To compute the coarse exponents, we need first to calculate 
the homogeneity volume $v_0$. We encounter here, of course, an old problem:
the determination of the scale of homogeneity of the universe. The SDSS data
have been employed for this purpose, with various results
\cite{SyL-Bar,Sar-Yad,Verev,CSF}. In fact, the ``scale of homogeneity'' is a 
loose concept and, as such, is bound to be defined in different ways, which produce different results, even if applied to the same data. A practical definition 
can be given in terms of the normalized second moment 
$\mu_2(v) = \langle \rho_v^2 \rangle = \M_2(v)/v$, where $\rho_v$ is the 
normalized coarse-grained density and
the last equality assumes that 
there are no empty cells in the lattice.
Homogeneity is defined by $\mu_2 = 1$, but this value is only approached asymptotically
for large $v$
(or as $v$ approaches the full sample volume).
The difference $\mu_2 - 1 = \langle (\rho_v-1)^2 \rangle$ measures the mass 
variance in the volume $v$ and can be used, in general,
to determine the regime of galaxy clustering \cite{I0}:
in the homogeneous regime, the 
coarse-grained density is Gaussian with a small or nearly vanishing variance.

The criterion for choosing $v_0$ in Refs.~\cite{MN,Bol} actually was that 
$\mu_2(v_0) = 1.1$, that is to say, a variance of 10\%.
This criterion is simple but is as arbitrary as any other, of course,
because one can as well demand smaller variances, say 5\%, 1\% or less, 
hence considerably increasing the value of $v_0$. Indeed, the claims that
homogeneity has not been found yet in galaxy catalogs are surely due to 
imposing too strict criteria for homogeneity.%
\footnote{In fact, what is claimed by some authors is that there are signs 
of inhomogeneity on very large scales \cite{SyL-Bar,Verev,CSF}.
However, certain signs of inhomogeneity that may look like structures can be observed
in a mass distribution with small mass-variance, for example, 
in a fluid in a critical state \cite{I0}.}
Without considering any specific property of the mass 
distribution, in addition to its having, on small scales, 
large values of $\mu_2$, that is to say, its being strongly 
non-Gaussian, one cannot propose a definite criterion and one is confined to
speculating about what is a sufficiently Gaussian distribution.
However, the scaling of $\M_q(v)$ and, in particular, of $\M_2(v)$ allows us 
to define a more precise scale of transition to homogeneity, 
namely, the scale
of crossover from the multifractal scaling with non-trivial values of $D_q$ to the homogeneous scaling with $D_q=3$ for all $q$.
It has been shown, in some cases, that this criterion roughly agrees with the 10\% mass 
variance criterion \cite{MN,Bol}.
Therefore, we also examine here the scaling of $\M_2(v)$. To be precise, 
we examine the crossover from the scaling $\mu_2 \sim v^{D_2/3-1}$ on middle 
scales to the exact value $\mu_2 =1$ on very large scales.
The value of $v_0$ obtained in this way is sufficient for our purpose, namely, 
for its use in Eq.~(\ref{ctauq}), and we do not need to consider subtle 
issues about the concept of homogeneity (see Ref.~\cite{I0}).

\subsection{Values of $v_0$ for our volume-limited samples}

\begin{figure}
\centering{
\includegraphics[width=7.1cm]{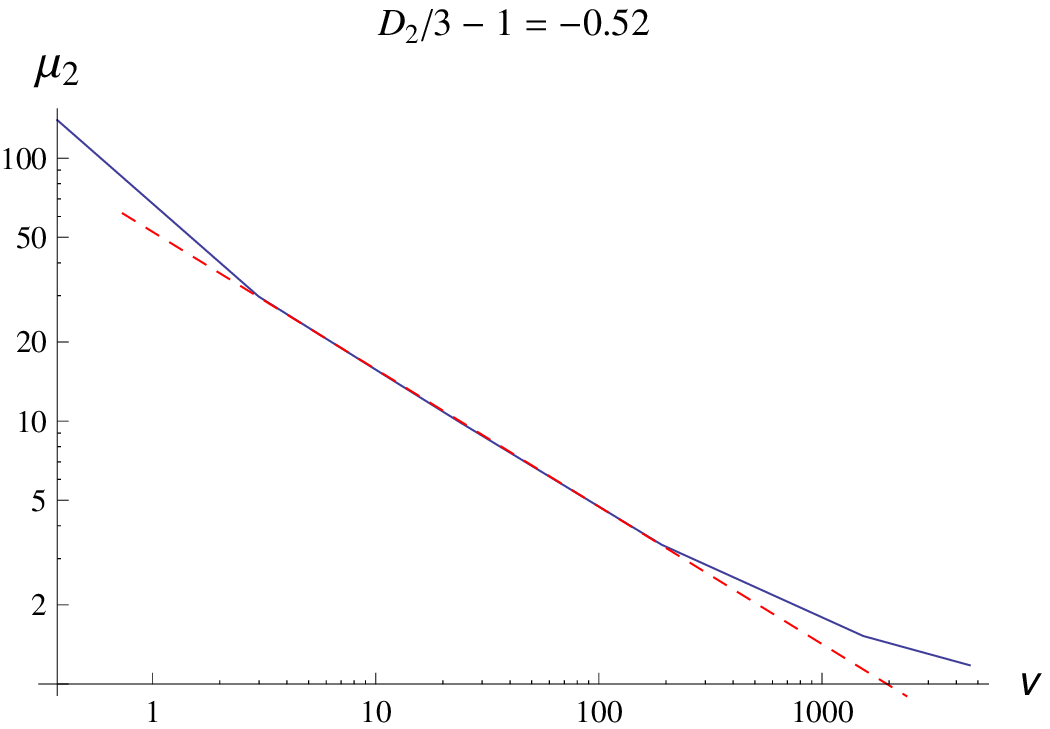}
\hspace{3mm}
\includegraphics[width=7.8cm]{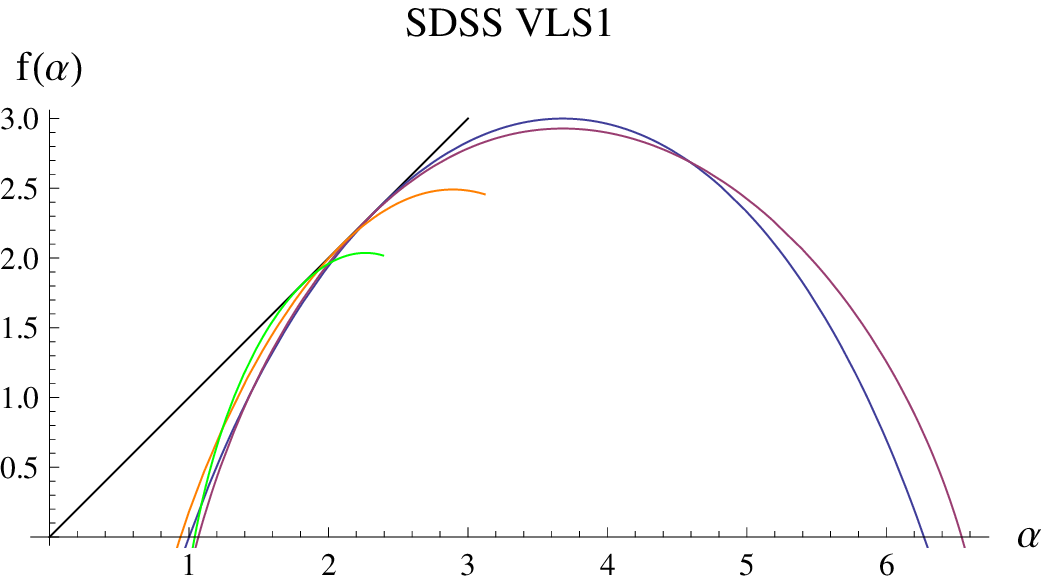}
}
\caption{Results of the analysis of sample VLS1: On the left-hand side, moment $\mu_2(v)$ 
($v$ in Mpc$^3/h^3$), with a fit of the scaling part $\propto v^{D_2/3-1}$. 
On the right-hand side, multifractal spectrum
for coarse-graining volumes 
$v = 2.99, \,23.9,\, 191.,\, 1.53\cdot10^{3}$ (Mpc/$h$)$^3$.}
\label{VLS1}
\end{figure}

The calculation of the scale of transition to homogeneity as 
the scale of crossover from a multifractal scaling 
to the homogeneous scaling in, for example, VLS1 
can be seen in the log-log plot of the corresponding $\mu_2(v)$, displayed in Fig.~\ref{VLS1}. This plot contains a 
fit of the scaling $\mu_2 \sim v^{D_2/3-1}$, 
in the interval $v\in [3, 190]$ Mpc$^3$/$h^3$,
with the result $D_2=1.44$. 
Of course, the fitting line and the consequent value 
of $D_2$ change if we change the interval of $v$, but one must choose
an interval that yields a good fit (with small errors).
The scale of crossover can be taken as the value of $v$ at the crossing 
of the fitting line (for the fractal scaling) and the 
line $\mu_2=1$ (for homogeneity), which yields $v_0=2000$ Mpc$^3$/$h^3$. 
However, 
the corresponding value of $\mu_2$ is 
somewhat high, that is to say, 
hardly compatible with a Gaussian distribution. So we 
take a larger value of $v_0$, such that the magnitude of 
$\mu_2-1$ is somewhat smaller. Notice that we can as well slightly increase 
the upper end of the interval of $v$ to fit, which will 
lower the absolute value of the slope 
and therefore increase the value of $v$ at the crossing point (and will also increase $D_2$).
Relying on these arguments, we take, for this sample, $v_0=4600$ Mpc$^3$/$h^3$
(the cell size in a $2\times 2 \times 2$ lattice, see Sect.~\ref{coarse}).

To $v_0=4600$ (Mpc/$h$)$^3$ corresponds 
the homogeneity length scale $4600^{1/3}$ Mpc/$h$ $=17$ Mpc/$h$, 
which is in reasonable agreement 
with the results from $N$-body simulations \cite{MN,Bol},
although it is smaller than other values for SDSS galaxies
\cite{SyL-Bar,Sar-Yad,Verev,CSF}. 
These values are obtained with different criteria, which are probably too strict, at least, for our purpose. Nevertheless, we find that the 
appropriate homogeneity scales for VLS2 and VLS3 are a little larger, reaching 25 Mpc/$h$ for VLS3. This value is practically equivalent to the 
value of 30 Mpc/$h$ obtained by Verevkin et al \cite{Verev}, also for the
SDSS-DR7 (although they counter that the uniform regime on larger scales still has some sort of inhomogeneities).

\subsection{Coarsening the volume-limited samples}
\label{coarse}

As explained in Sect.~\ref{MF}, our multifractal analysis is carried out in 
lattices of equal-volume cells that are formed by a particular Cartesian product.
This product results from multiplying
an angular-coordinate lattice, 
with the intervals of $sl$ and $f$ divided 
into equal subintervals, 
by 
the interval of $r$ divided into subintervals with constant increment of $r^3$. 
Additionally, we want the resulting cells to be reasonable regular, with aspect ratios not very 
different from one. 

For each sample, we must prepare a set of diminishing meshes.
To achieve good aspect ratios, we need to
adapt the set of meshes to the particular interval of $r$ of each sample. 
For the sake of computational simplicity, 
once the initial coarse mesh is chosen for a sample, 
we generate a sequence of finer meshes 
by using binary division of subintervals.
Given the ranges of $sl$ and $f$, 
respectively, $1.55$ and $1.22$, 
an initial $4\times 3$ angular lattice 
generates angular cells with aspect ratio close to one (see the right part of Fig.~\ref{slice}). The interval of $r$ depends on the sample, but
the shape of the spherical sector for a sample is 
only determined by the ratio of the upper to the lower limits of $r$. 
This ratio is almost the same for VLS2 and VLS3, and it is such that for all the three
samples the length of the radial interval is smaller than the length of the angular intervals.
We find that initial coarse meshes of
$4\times 3 \times 2$ and $4\times 3 \times 1$ yield 
acceptable aspect ratios for, respectively, VLS1 and VLS2 (or VLS3).
The volume of the cells in the initial 24-cell lattice for VLS1 is actually 
somewhat small, 
so we add in this case a coarser mesh, 
namely, a $2\times 2 \times 2$ mesh, to probe larger scales and, in particular, 
to find the transition to homogeneity.

\subsection{Multifractal spectrum}
\label{MFspec}

Several coarse multifractal spectra of VLS1, computed with Eq.~(\ref{ctauq})
and 
$v_0=4600$ Mpc$^3$/$h^3$,
are displayed in Fig.~\ref{VLS1}. The agreement with the spectrum 
obtained from $N$-body simulations (Sect.~\ref{MF-Bolshoi} and 
Fig.~\ref{Bolshoi_spec}) is very convincing in 
the zone $\a<3$, corresponding to singularities and therefore galaxy clusters. Indeed, 
the dimension of the mass concentrate, given by the points of tangency to the diagonal,
is between 2 and 2.6 and is probably about 2.4;
and the strongest singularities also have $\a \simeq 1$.
For $\a \gtrsim 3$, we have convergence of only two coarse multifractal spectra, but 
they show anyhow that the maximum value of $f(\a)$, 
namely, the box-counting dimension of the distribution's support, is very close to 3.
However, the fact that all (or almost all) cells are non-empty 
for just two scales, which are not well below $v_0$, is not a fully 
convincing proof of non-lacunarity.
Anyway, the concordance between this galaxy sample and the results of $N$-body simulations is remarkable, because the mass resolution of galaxies is generally much worse and, 
furthermore, galaxy data are subject to considerable errors (see below). 
Unfortunately, the concordance breaks down
for $\a > 4$ and, in particular, the largest value
of $\a$ is notably larger in Fig.~\ref{VLS1} than in 
Fig.~\ref{Bolshoi_spec}.

\begin{figure}
\centering{
\includegraphics[width=7.5cm]{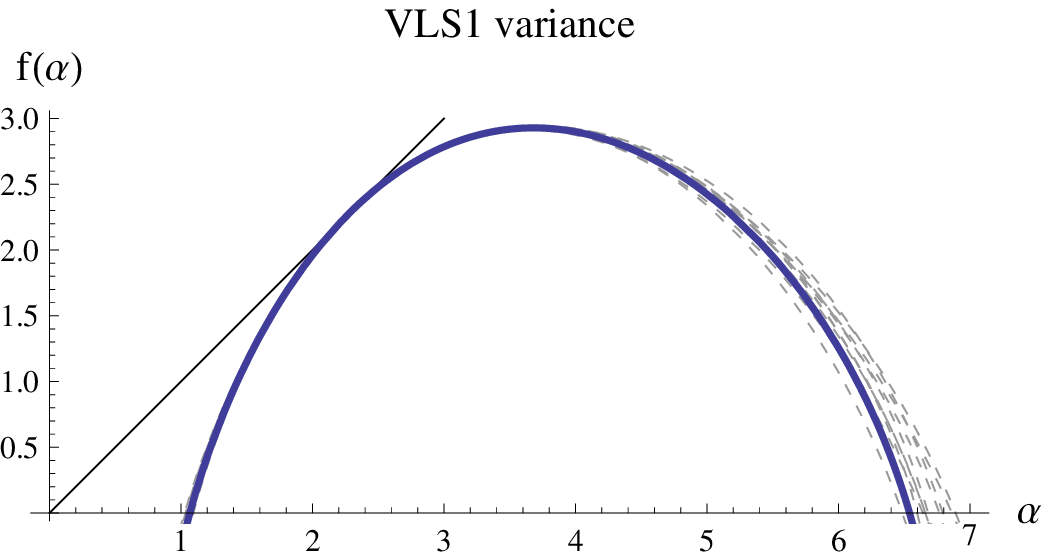}
\hspace{3mm}
\includegraphics[width=7.3cm]{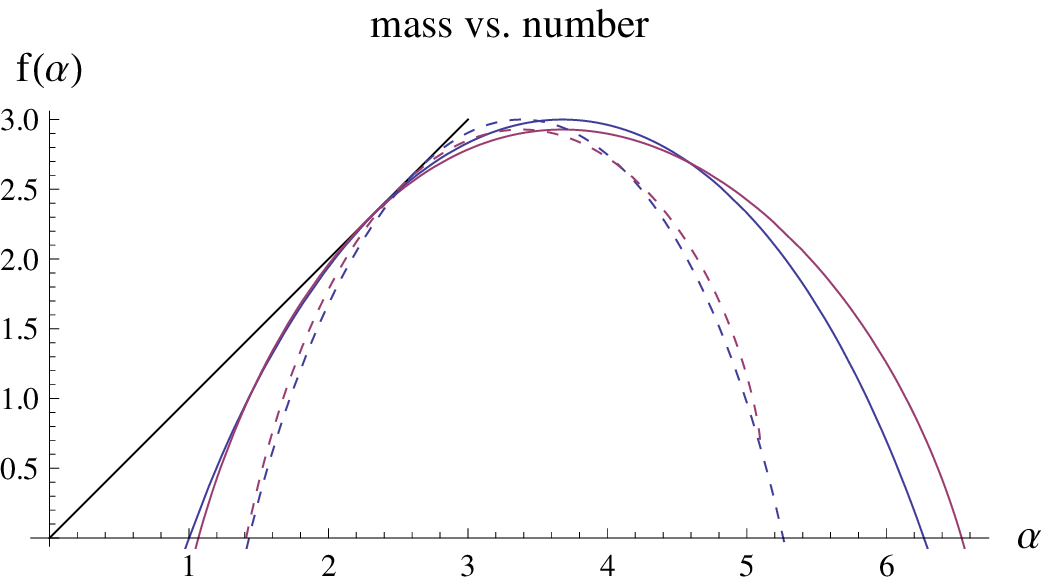}
}
\caption{Errors in multifractal spectra: (Left) Variance in the spectrum of VLS1 at 
$v = 191$ (Mpc/$h$)$^3$ due to errors in galaxy positions and masses (ten variant spectra, 
in dashed lines).
(Right) Effect of suppressing galaxy masses in VLS1 at
$v = 191.,\, 1.53\cdot10^{3}$ (Mpc/$h$)$^3$ (in dashed lines).}
\label{VLS1_error}
\end{figure}

These results are supported by the estimation of errors in the VLS1 coarse spectra, due to 
errors in galaxy positions and masses. The errors in angular and radial coordinates have 
different origin: the uncertainty in angular positions is mainly due to the size of galaxies, whereas the uncertainty in distance is mainly due to the uncertainty in the Hubble law 
caused by peculiar velocities. The size of the type of galaxies in VLS1 can be calculated 
in terms of galaxy mass according to $R(\mathrm{kpc}) = 0.1(M/M_\odot)^{0.14}$ \cite{Shen}. 
This yields a maximum radius $< 2$ kpc, which is negligible in comparison with the sizes of 
the coarse-graining cells, larger than 1 Mpc/$h$ (the error is just a bit larger than 
the error in the Bolshoi simulation, Sect.~\ref{MF-Bolshoi}). As regards radial coordinates,
peculiar velocities actually destroy 
the Hubble flow on small scales and, hence, the determination of distance by redshift. 
However, the local Hubble flow is ``cold'' and the dispersion of peculiar velocities
is as low as 30 km/s \cite{Kara}. To be on the safe side, we take a dispersion of 50 km/s. 
Therefore, we assume, for the error in distance, that 
redshifts have normal (Gaussian) errors with dispersion $\sigma_z = 1/6000$.
Notice that this error is certainly non-negligible at the lower cut in redshift,
$z=0.003$.

The error in stellar mass estimates is such that $\log_{10}\!M \pm 0.3$ spans the 
95\% confidence interval \cite[Fig.~15]{Kauff}.  
Given that a $2\sigma$ Gaussian deviation interval comprises 95\% of probability,
we can assume for $M$ a lognormal distribution with 
$\sigma_M = (\ln 10)\,0.3/2=0.3454$ (using natural logarithms). This distribution is 
quite skewed, because the dispersion is considerable: indeed, the 95\% confidence interval
amounts to a factor of 4 in mass.  Notice that the uncertainty in mass is much
larger than the uncertainty in position.

Once we have the $z$ and $M$ distributions, we can generate a number of random alternatives 
to our initial sample, before the construction of the VL samples. 
From those alternative initial samples, 
we construct the corresponding variants of VLS1, following the same procedure 
followed for VLS1 itself. Hence, we compute the corresponding coarse multifractal spectra.
We have done so for ten alternative initial samples, focusing on the 
most relevant VLS1 multifractal spectrum, namely, the spectrum of VLS1 at 
$v = 191$ (Mpc/$h$)$^3$, which is within the scaling range yet  
is reasonable in the void zone ($\a>3$), as seen in Fig.~\ref{VLS1}.
The variance in this coarse multifractal spectrum is displayed in 
the left-hand side of Fig.~\ref{VLS1_error}. The graph shows that 
the error is negligible in the cluster zone ($\a<3$), and even so up to the maximum 
of $f(\a)$. The error grows for $\a>4$ and is 
considerable at the end of the spectrum, in accord with 
the insufficient mass resolution in that zone, as manifested by the convergence of only 
two coarse spectra.

One further ``error check'' is carried out by considering 
the effect of making all the galaxy masses equal. 
Such a drastic simplification is beyond any magnitude of error in stellar mass estimation 
but is relevant for a comparison with the approach that only considers 
the number of galaxies. Employing two relevant coarse multifractal spectra of VLS1, namely, 
the ones for $v = 191$ and $1.53\cdot10^{3}$ (Mpc/$h$)$^3$, we see in 
the right-hand side of Fig.~\ref{VLS1_error} the effect of suppressing
mass information. Remarkably, the range of $\a$ shrinks, 
more at the right end, belonging to voids, than at the left end, belonging to clusters. 
But the effect is notable in both cases and actually is more important if
$\a < 3$, because this is the more reliable part of the spectrum. In fact, by 
suppressing mass information, we ruin the concordance with 
the spectrum obtained from $N$-body simulations in that zone.
The magnitude of the effect of suppressing masses is undoubtedly due to 
the broad range of masses (see Table \ref{VLStable}).

\begin{figure}
\centering{
\includegraphics[width=7.4cm]{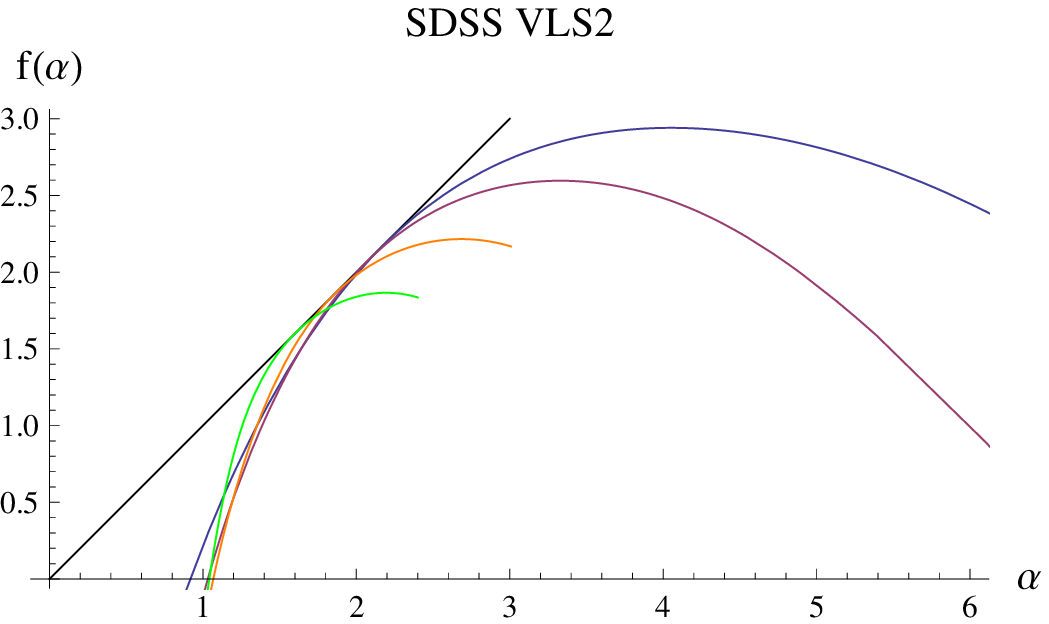}
\hspace{3mm}
\includegraphics[width=7.4cm]{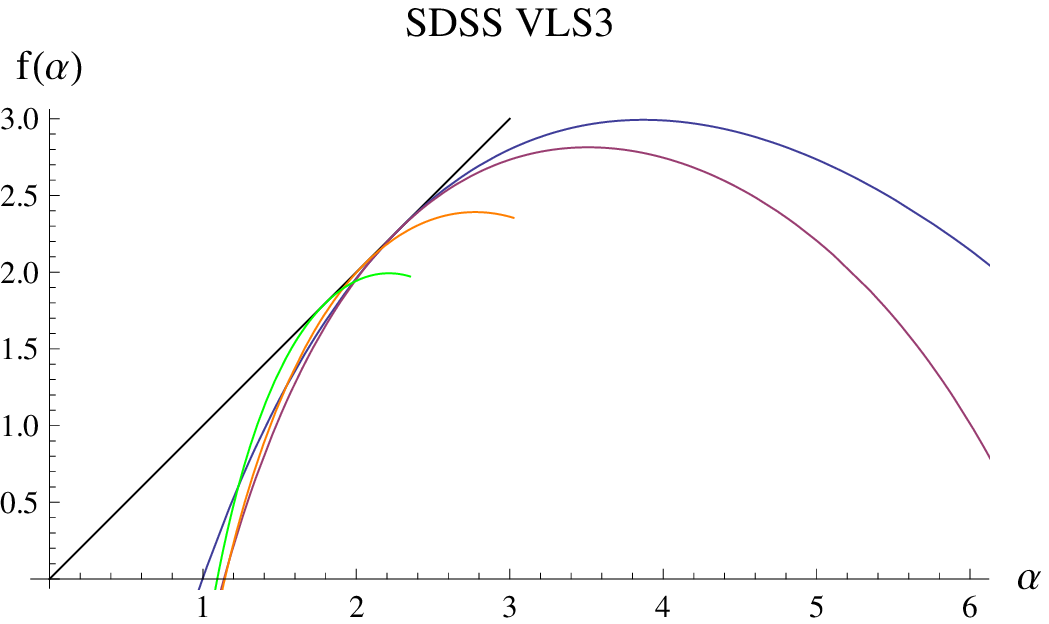}
}
\caption{Results of the analysis of samples VLS2 and VLS3.}
\label{VLS_2-3}
\end{figure}

We now proceed to the analysis of deeper VL samples, namely, VLS2 and VLS3.
It does not provide new 
information: the results for clusters are nearly the same as before but
the results for voids are definitely worse (Fig.~\ref{VLS_2-3}).
In fact, the absence of any convergence of coarse multifractal spectra 
in the zone $\a>3$ shows that these deeper VL samples are useless for the 
study of voids: the voids in them are due to undersampling. 
Coleman and Pietronero \cite{Cole-Pietro}, in their multifractal analysis of 
the CfA catalog, already noticed that 
$\tau(q)$ cannot be obtained for negative $q$ because $q<0$ corresponds to
the lowest masses, which are not well represented. 
Indeed, $\tau(q)$ for $q<0$ yields the part of $f(\a)$ beyond its maximum. 
At any rate, the part of $f(\a)$ up to its maximum that
Coleman and Pietronero \cite{Cole-Pietro} calculate does not agree with 
our results: they obtain that $\a_\mathrm{min}=0.65$ and that the maximum 
value of $f$ is equal to 1.5, in contrast with our values $\a_\mathrm{min}=1$
and maximum of $f$ equal to 3. The multifractal analyses of galactic catalogs by other authors were based on the galaxy number density but obtained results similar to Coleman and Pietronero's, 
that is to say, obtained that $\a_\mathrm{min}$ is smaller 
than one and that the maximum of $f$ is quite smaller than three.
The use of better data and the reasonable agreement with 
the results of $N$-body simulations in our analysis make it more reliable.

Let us compare our multifractal analysis with the recent 
analysis of the SDSS-DR7 by 
Chac\'on-Cardona {\em et al} \cite{CSF}. This analysis is made in terms 
of R\'enyi dimensions $D_q$, which provide equivalent information to 
the multifractal spectrum $f(\a)$, from a mathematical viewpoint (Sect.~\ref{MF}).
However, Chac\'on-Cardona {\em et al} select
deep VL samples, consider a distribution of dark matter halos 
associated to SSDS-DR7 galaxies instead of the galaxies themselves, 
and do not take into account the galaxy masses.
Moreover, their definition of $D_q$ as a derivative with respect to scale 
is different from the standard definition of fractal dimension adopted here, 
in terms of the limit for vanishing scale 
\cite{Falcon,Harte}.
In consequence, it is difficult to compare directly the present analysis with 
the analysis by Chac\'on-Cardona {\em et al} \cite{CSF}. 
Naturally, the first problem is to compare their results for $D_q$ with the present results for $f(\a)$. A few partial comparisons are simple to make: for example,
the value of the maximum of $f(\a)$ must be equal to $D_0$. 
However, while we have found, with confidence, 
that the maximum of $f(\a)$ is 3, no definite value of $D_0$
can be deduced from \cite[Fig.~4]{CSF}.

At any rate, we believe that the knowledge 
of the multifractal spectrum $f(\a)$ of 
the large scale stellar-mass distribution
is more directly useful than the knowledge of R\'enyi dimensions, 
in general, because it allows us to deduce several consequences, which we discuss next. 

\section{Conclusions and discussion}
\label{discuss}

We have calculated the multifractal spectrum of the large scale 
stellar-mass distribution, employing the SDSS-DR7, in an 
effort to determine the geometry of the baryonic cosmic web and see
how it relates to the geometry of the dark matter cosmic web.
Of course, the 
stellar mass is only a fraction of the total baryonic mass and, furthermore, the 
SDSS data do not contain all the stellar mass, because of the 
cuts in apparent magnitudes of galaxies (and the additional cuts in VL samples). 
Nevertheless, the information 
obtained is representative.
A different type of information on the distribution of baryonic mass
is provided by $N$-body simulations containing baryonic gas. 
From one of them, we can conclude 
that the fractal geometry of the distributions of gas and dark matter 
is the same. Assuming this identity, 
the question that we are addressing 
is whether or not the fractal geometry of the distribution of the visible stellar mass coincides with that already known common geometry.

Unfortunately, even the rich SDSS data are insufficient for fully 
determining the multifractal geometry of the stellar mass distribution.
At any rate, 
we can assert the overall consistency of the stellar mass multifractal spectrum,
namely, the internal consistency of our multifractal analysis of 
the SDSS data and, furthermore, its consistency with the multifractal 
analysis of LCDM $N$-body simulations. While
the internal consistency is clear in the case of 
mass concentrations (clusters), it is however questionable in the case of 
mass depletions (voids), since we only have convergence, at best, of two coarse multifractal spectra (the minimum number to speak of convergence) and only for one sample.
Therefore, 
we discuss the cluster ($\a < 3$) and void ($\a > 3$) cases
separately, beginning with the former. 

The value $\a_\mathrm{min} = 1$, common to our analysis of the SDSS-DR7 and 
$N$-body simulations, corresponds to the singular isothermal sphere profile, 
the standard profile of the outskirts of individual galaxies,
and also corresponds to a filament, a basic element of the cosmic web.
More in general, it is a natural lower limit, because the 
gravitational potential diverges at a point 
on which mass concentrates with $\a < 1$. Such a
mass concentration would not only have a divergent mass density, which may 
not be physically forbidden, but would also involve a divergent (negative) gravitational energy, 
which is certainly forbidden. 
Of course, the gravitational energy would not actually become infinite, 
because the cosmic-web mass distribution is not valid down to
infinitely small scales and the scaling law $m(r) \sim r^\a$ must change
as $r\ra 0$, but mass concentrations with $\a < 1$ are anyhow linked to excessive 
energy dissipation and, therefore, must be unlikely to appear.
Let us remark that some mass concentrations with $\a <1$ do appear in our analysis, 
but they have $f(\a)<0$ and are not displayed in Fig.~\ref{VLS1} or Fig.~\ref{VLS_2-3}. 
Moreover, $\a <1$ values with $f(\a)<0$ also appear in the analysis of 
$N$-body simulations \cite{MN}. However, mass concentrations with $f(\a)<0$ tend to
disappear as $v\ra 0$, as explained in Sect.~\ref{MF}, and this is indeed what we observe.

In this regard, let us notice that the adhesion model predicts 
that {\em knots} are widespread in the cosmic-web, and these knots
are points with finite mass, that is to say, singularities of maximum strength, $\a=0$.
They do not seem to be present in the real cosmic web. 
The reason for this discrepancy is, of course, that 
the Zeldovich approximation can only describe 
gravitational dynamics
on the larger scales and is unable to describe gravitational collapse with 
strong energy dissipation.

Singularities with $\a$ close to one are very significant 
energy-wise, but the total mass that they contain is insignificant.
The bulk of the mass concentrates on a set of singularities with dimension 
$\a_1 = f(\a_1) = D_1 \simeq 2.4$. 
This is a remarkably high value, especially, when we compare it with the results 
of older multifractal analyses of the galaxy distribution (with or without galaxy masses), 
which obtain a {\em maximum} of $f(\a)$ that is about 2: 
but the dimension of the mass concentrate must be lower than the maximum dimension, of course. 
As regards the cosmic-web morphology, one could be tempted to 
conclude that such a high dimension of the mass concentrate 
favors cosmic sheets over filaments. However, such conclusion would not be 
warranted at all, because the fractal dimension does not directly give information on morphology 
and one should employ instead the {\em topological dimension} and measures of {\em texture} 
\cite{Mandel}.

Regarding void regions, formed by points with $\a > 3$, 
the main conclusion is that the maximum value of $f(\a)$, 
which gives 
the (box-counting) dimension of the support of the distribution, is very close to 3
(and is at $\a \simeq 3.5$). 
The dimension 3 corresponds to non-fractal support and suggests that the cosmic web is a
non-lacunar multifractal.
To confirm it, a specific study of voids is necessary. 
From the study of voids in Ref.~\cite{voids} and from 
the multifractal analysis of $N$-body simulations,
it can be concluded with confidence that 
the geometry of the dark matter or the baryonic gas is non-lacunar,  
while the present analysis of 
the stellar mass distribution is not as conclusive.
Voids can be
perceived in the galaxy distribution (e.g., in the right-hand side of 
Fig.~\ref{slice}) but they are, presumably, an effect of undersampling. 
This effect can combine with the existence of regions with very low baryonic density
and therefore very few stars. 
It is to be remarked that 
the cosmic web generated by the adhesion model, although somewhat different 
from the real cosmic web, is also an example of
the peculiar geometry of non-lacunar fractals 
\cite{voids}.

For strong mass depletions, with $\a > 4$, the result of our analysis of the SDSS
differs from the result of analyses of $N$-body simulations. 
The latter shows a 
rather sharp decline of the dimension of point sets with $\a > 4$ whereas
the decline shown by the SDSS analysis is more progressive. We must caution that 
the SDSS data are hardly sufficient to be confident about this conclusion.
In this regard, we must wait for more information,
from future surveys able to detect much fainter galaxies. 
Other types of data can also be employed; for example, 
data on the intergalactic matter.

Finally, we can compare the value $D_2=1.44$, obtained by
fitting the scaling $\mu_2 \sim v^{D_2/3-1}$ in VLS1, with other values of the correlation dimension. Sylos Labini {\em et al} \cite{SyL-Bar} and
Verevkin {\em et al} \cite{Verev} obtain $D_2=2$ (or higher) from the SDSS data, 
employing the DR4 and DR7, respectively. However, smaller values, obtained 
from various samples, appear in the literature
\cite{Jones-RMP}. We must caution that the often calculated and discussed
correlation dimension 
is always the one that corresponds to the 
galaxy position correlation function, whereas our value of $D_2$ corresponds instead to the correlation function of the stellar mass distribution.

\acknowledgments
I thank C.A.\ Chac\'on-Cardona for correspondence and for the SDSS-DR7 file.

\end{document}